\documentclass[apj]{emulateapj}

\usepackage{graphicx}
\usepackage{microtype}
\usepackage{natbib}
\bibpunct{(}{)}{;}{a}{}{,} % to follow the A&A style
\usepackage{amssymb}
\usepackage{hyperref}  % PDF hyperlinks, with coloured links
\def\link_col{blue}
\hypersetup{colorlinks,citecolor=\link_col,filecolor=\link_col,linkcolor=\link_col,urlcolor=\link_col}
\hypersetup{linkcolor=red,citecolor=blue,filecolor=magenta,urlcolor=blue} % coloured links
\usepackage{xspace}
\usepackage{url}
\usepackage{wasysym}

\def \cmcube{\mbox{cm$^{-3}$}\xspace}
\def \cmsqr{\mbox{cm$^{-2}$}\xspace}
\def \pers{s$^{-1}$\xspace}

\def \nhthree{\mbox{NH$_{3}$}\xspace}
\def \C34S{\mbox{C$^{34}$S}\xspace}
\def \13CS{\mbox{$^{13}$CS}\xspace}
\def \msun{\mbox{M$_{\odot}$}\xspace}

\def \water{H$_2$0\xspace}
\def \source{G0.216+0.016\xspace}

\def\mug{$\mu$G\xspace}

\defcitealias{Jones2008}{J08}
\defcitealias{Protheroe2008}{P08}

\slugcomment{Accepted to the Astrophysical Journal Letters}

\shorttitle{Prospects for Detection of Synchrotron Emission from Secondaries Near Dense Dust Clumps}
\shortauthors{Jones, D. I.}

\begin{document}

\title{Prospects for Detection of Synchrotron Emission from Secondary Electrons and Positrons in Starless Cores: Application to \source}

\author{Jones, D. I.}
\affil{Department of Astrophysics/IMAPP, Radboud University, Heijendaalseweg 135, 6525 AJ Nijmegen, The Netherlands.}
\email{d.jones@astro.ru.nl}

\begin{abstract}
We investigate the diffusion of cosmic rays into molecular cloud complexes.
Using the cosmic-ray diffusion formalism of Protheroe, et~al. (2008), we examine how cosmic rays diffuse into clouds exhibiting different density structures, including a smoothed step-function, as well as Gaussian and inverse-$r$ density distributions, which are well known to trace the structure of star-forming regions.
These density distributions were modelled as an approximation to the Galactic centre cloud \source, a recently-discovered massive dust clump that exhibits limited signs of massive star formation and thus may be the best region in the Galaxy to observe synchrotron emission from secondary electrons and positrons.
Examination of the resulting synchrotron emission, produced by the interaction of cosmic ray protons interacting with ambient molecular matter producing secondary electrons and positrons reveals that, due to projection effects, limb-brightened morphology results in all cases.
However, we find that the Gaussian and inverse-$r$ density distributions show much broader flux density distributions than step-function distributions.
Significantly, some of the compact (compared to the $2.2''$ resolution, 5.3~GHz JVLA observations) sources show non-thermal emission, which may potentially be explained by the density structure and the lack of diffusion of cosmic rays into the cloud.
We find that we can match the 5.3 and 20~GHz flux densities of the non-thermal source JVLA~1 and 6 from Rodr{\'{\i}}guez \& Zapata (2014) with a local cosmic ray flux density, a diffusion coefficient suppression factor of $\chi=0.1-0.01$ for a coefficient of $3\times10^{27}$~\cmsqr~\pers, and a magnetic field strength of 470~\mug.
\end{abstract}

\maketitle

\section{Introduction}
The question of the penetration of cosmic rays (CRs) into molecular clouds and any resulting emission is an important question in high-energy astrophysics.
\citet{Gabici2007} considered CR diffusion (and resulting CR proton interactions) into a typical molecular cloud of $n_{\rm{\tiny H}_2}=300$~\cmcube, $B=10$~\mug and CR diffusion coefficients typical for the Galactic disk.
They found that GeV to TeV energy CR protons would indeed penetrate to the centre of such a cloud, and thus clouds should be a target for gamma-ray observations.
At GHz radio frequencies, this hypothesis has been tested several times.
Firstly, \citep[hereafter J08]{Jones2008} searched two massive, isolated molecular clouds for evidence of synchrotron emission on the basis that the same protons that produce gamma rays (through inelastic $pp$ collisions and subsequent neutral pion decay) will also produce MeV--GeV secondary electrons and positrons\footnote{Hereafter, for brevity, we refer to any secondaries produced as simply electrons} (through the decay of the charged pions produced concomitantly with the neutral pions) that will radiate at GHz frequencies.
However, the confusion of any possible non-thermal emission with optically thick and thin thermal emission produced by star-formation processes, made any detection all but impossible. 
Furthermore, \citet[hereafter P08]{Protheroe2008} and \citet{Jones2011} analysed radio emission from the Sagittarius~B2 (Sgr~B2) giant molecular cloud for such synchrotron emission because of the enhanced CR flux density thought to be present in the central regions of our Galaxy.
The main results of these studies are that the diffusion of CRs are severely limited into dense cores of molecular clouds, and that clouds with a Gaussian density structure (such as suggested by \nhthree(1,1) studies by \citealt{Ott2014}) will show a ``limb-brightened'' morphology.
The outcome of these investigations is that it is vital for unambiguous detection of synchrotron emission at GHz frequencies that the target clouds that, whilst being massive ($\gtrsim10^4$~\msun) and dense ($\gtrsim10^3$~\cmcube), also do not show signs of star formation, since the thermal radiation that forming stars quickly produce, will swamp any ability for detection of non-thermal emission.
These considerations leave precious few clouds as potential targets.
Moreover, as \citet{Ginsburg2012} found no starless, massive dust clumps in the first Galactic quadrant -- implying a very short timescale for massive star formation (viz. $\lesssim0.5$~Myr) -- the window to observe synchrotron emission from secondary electrons in massive clumps is thus also likely to be short.

The Galactic centre cloud, \source, was recently shown -- at least in terms of massive star, and possibly globular cluster, formation -- to be a very important object \citep{Longmore2012}.
Though it was discovered 20 years ago \citep{Lis1994}, it is remarkable for such a massive ($\sim10^5$~\msun), dense ($\sim2.0\times10^4$~\cmcube) clump to exhibit a weak water maser as the sole sign of massive star formation.
Given that a radio study of \source \citep{Rodriguez2013} at 5.3 and 20.9~GHz with the Karl G. Jansky VLA (JVLA) showed only a small number ($<7$) of dense radio continuum clumps, this suggests that this cloud is the best chance in the Galaxy to observe synchrotron emission from secondary electrons.
Some important information about \source is summarised in Table~\ref{table:comp}.
That three of the sources found by \citet{Rodriguez2013} exhibit non-thermal spectra thus compelled us to model the structure of this cloud in order to investigate whether, given the high mass of the cloud, synchrotron emission from secondary electrons, together with the expected exclusion of the CRs from the dense parts of the cloud, could explain the observed non-thermal emission.

\begin{deluxetable}{llll}
\tablewidth{0pt}
\tablecaption{Mass, density and star-formation signposts associated with \source.\label{table:comp}}
\tablehead{
\colhead{Mass} & \colhead{$\langle n_{H_2}\rangle$ } & \colhead{maser(s)?} & \colhead{HII region(s)?} 
} 
\startdata
$2\times10^5$ & $2\times10^4$ & \water & 3 \\
\enddata
\end{deluxetable}

\section{Predicted Synchrotron Emissivity and CR Diffusion}\label{sec:model}
\subsection{Synchrotron Emissivity due to Secondary Electrons}
As outlined in \citetalias{Jones2008}, the synchrotron emissivity, $j_\nu$, due to secondary electrons, in units of $erg$~\cmcube~$s^{-1}$~sr$^{-1}$~Hz$^{-1}$, at frequency, $\nu$, is obtained by appropriate integration over the ambient electron number density spectrum:
\begin{equation}\label{eq:synch}
	j_\nu({\mathbf r})=\frac{\sqrt{3}e^3}{4\pi m_e c^2}\left(\frac{B_\perp}{1\rm{~G}}\right)\int^\infty_{m_e c^2}F(\nu/\nu_c)n(E, {\mathbf r})d E,
\end{equation}
where $e=4.8\times10^{-10}$~e.s.u. is the charge of the electron and $m_e c^2=8.18\times10^{-7}$~erg, $B_\perp$ is the  magnetic field strength in Gauss, and $F(\nu/\nu_c)$ is the first synchrotron function evaluated as a function of the critical frequency; $\nu_c=4.19\times10^6(E/m_e c^2)^2(B_\perp/1{\rm~G})$~Hz.
Here, $n(E, {\bf r})dE$ is the ambient spectrum of secondary particles, we have taken the same losses (ionisation, bremsstrahlung and synchrotron) into account as in \citetalias{Jones2008}:
\begin{equation}
n(E,{\mathbf r}) = \frac{\int^\infty_Eq_\pm(E,{\mathbf r})dE}{dE/dt},
\end{equation}
for an appropriate production spectrum of secondary electrons, $q_\pm(E,{\bf r})$, for which we use the spectrum of \citet{Kamae2006}, normalised to the proton spectrum found by the PAMELA experiment \citep{Adriani2011}.

\begin{figure*}
  \centering
 \includegraphics[width=0.5\textwidth]{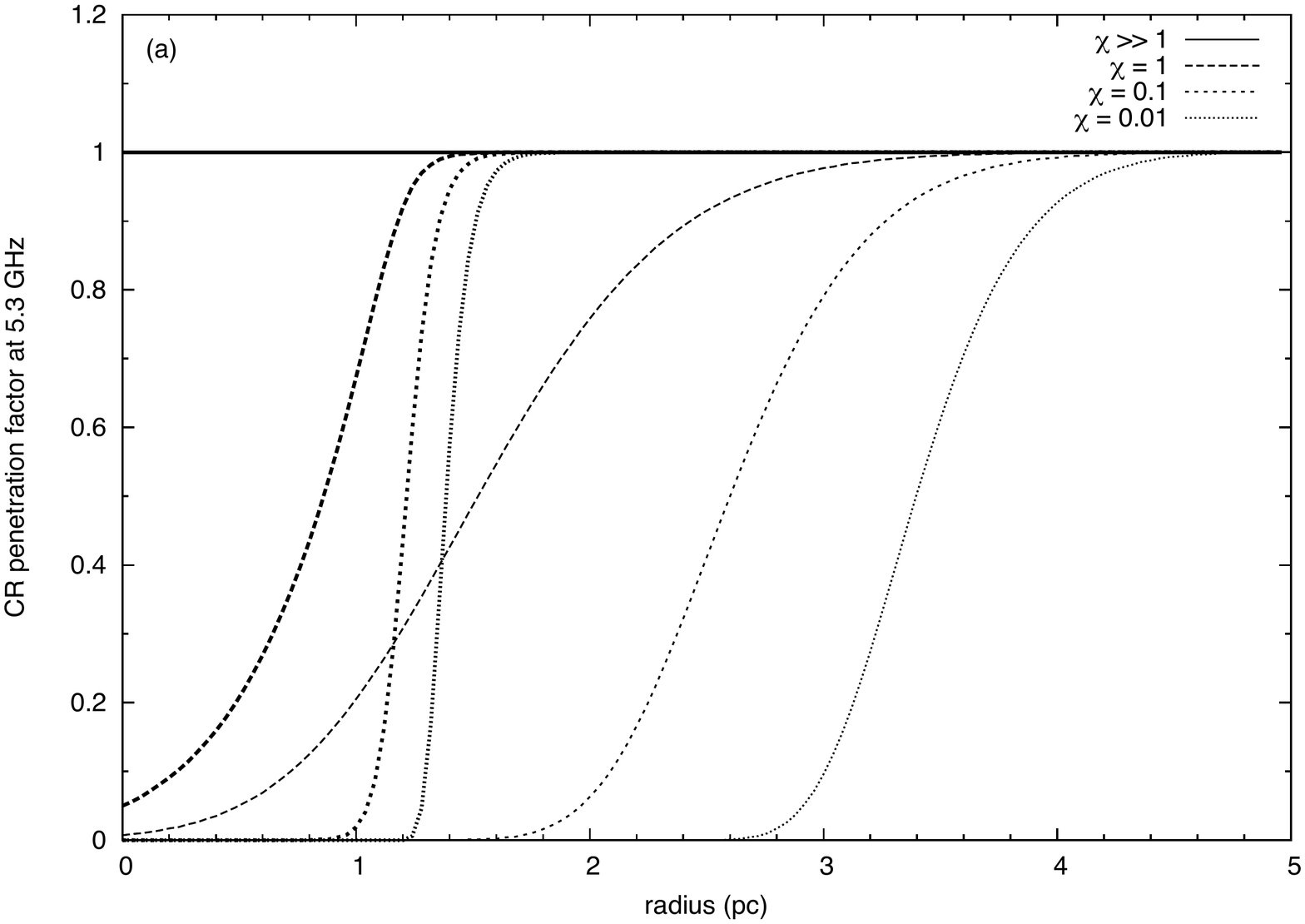}\includegraphics[width=0.5\textwidth]{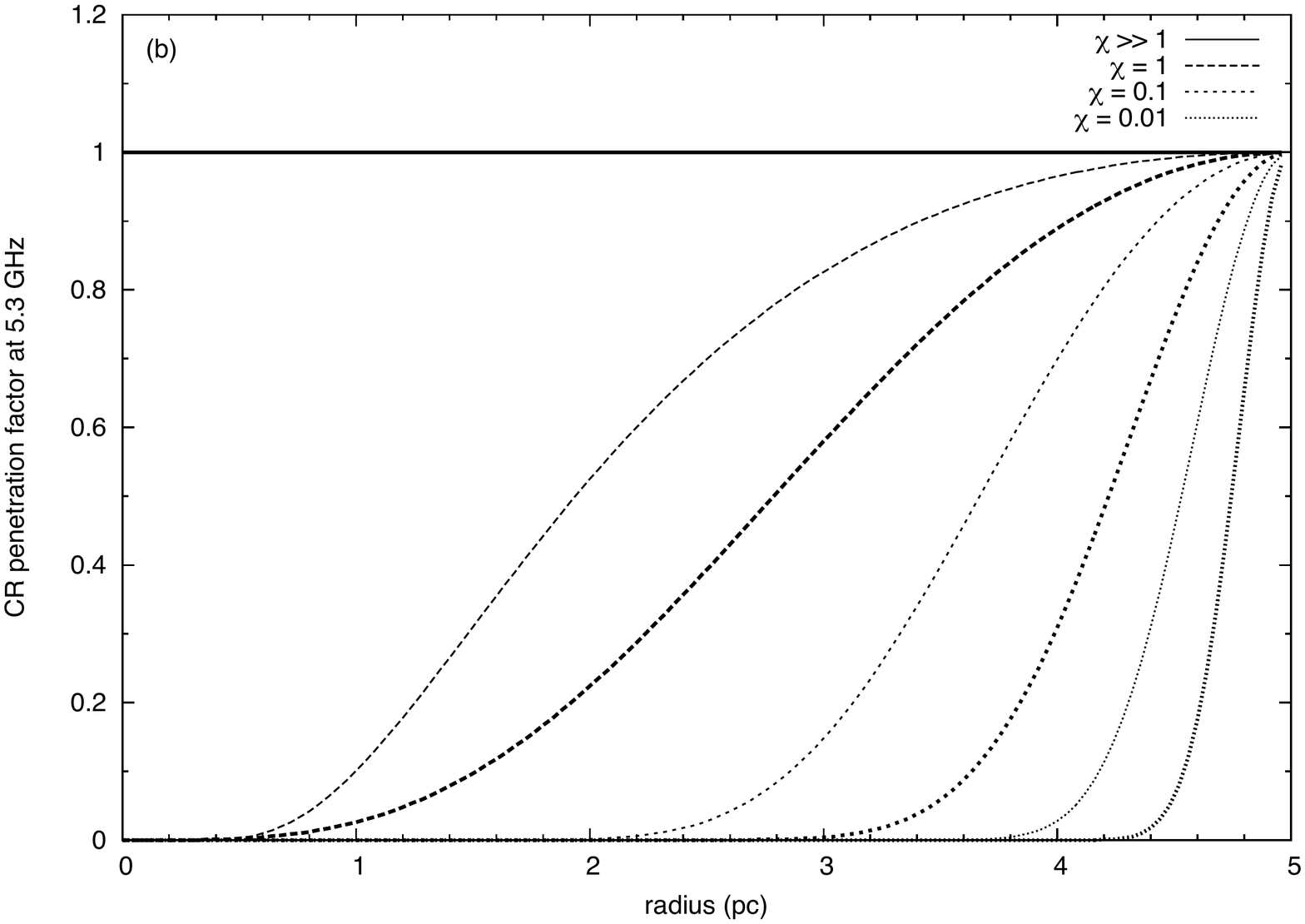}
  \caption{Penetration factor, $e^{-\tau_\star}$, of CRs into a molecular cloud with a magnetic field strength of 600~\mug, peak density of $2\times10^4$~\cmcube and a mass of $7\times10^5$~\msun.
  Four different density structures are considered, (a) a Gaussian of standard deviation $\sigma=2.95$~pc (thin lines) and step-function (thick lines; see text for more details) of the same width, and (b) and $1/r^2$ (thin lines) and $1/r^{1/1}$ (thick lines), where the radius, $r$ is normalised to the same characteristic width as in (a).
  Each line is for a different CR diffusion suppression factor, $\chi$, as labelled in the figure.
  }
  \label{fig:CRpen}
\end{figure*}

\subsection{Density Distribution of the Cloud}\label{sec:density}
As will be shown below, an estimate of the CR distribution within a cloud is critically dependant on how the density structure of the cloud is modelled.
\citetalias{Protheroe2008} showed that the large-scale distribution of molecular material in Sgr~B2 can be modelled well by a 2-dimensional Gaussian structure with standard deviation $\sigma=2.75$~pc.
On the other hand, as discussed in \citet{Johnston2014}, probability density functions (PDFs) of either the volume or column density have been used in an attempt to elucidate the structures of molecular clouds and to investigate the various physical processes contributing to this overall structure.
It has been shown that PDFs of active Galactic star forming regions actively forming stars show a power-law tail, in addition to a log-normal PDF (\citet{Johnston2014} and references therein).
In virial mass modelling, the density profile is modelled as $\rho({\bf r})\propto r^p$.
Studies have shown that the median power for a sample of star-forming regions is $p=-1.8$ \citep{Mueller2002}, similar to the $p=-1.1,-2.0$ models discussed here.
However, given that \source shows no sign of such a power-law tail \citep{Johnston2014}, and that Figure~9 of \citet{Rathborne2014} shows a ``centrally condensed'' density structure for \source, we have modelled such a structure using a smoothed step-function with functional form:
\begin{equation}
f(r) =\frac{ae^{cr} + be^{dr}}{e^{cd} + e^{dr}}.
\end{equation}
This function produces a step function from $a$ to $b$ at a value of $c=2.95$~pc, with $d$ controlling the ``steepness'' of the cutoff, with smaller values giving a sharper cutoff (here, we use the somewhat-arbitrarily chosen $d=6$, though we find that modest changes in $d$ do not significantly effect the overall structure).
However, in order to more fully explore the density-structure parameter space, we have also modelled a Gaussian of standard deviation $\sigma=2.95$~pc as well as inverse-$r$ density distributions ($1/r^2$ and $1/r^{1.1}$).
We have normalised the step function and inverse-$r$ density distributions to the peak density of $2\times10^4$~\cmcube obtained by \citet{Rathborne2014}.

\subsection{CR Diffusion into Molecular Clouds}\label{sec:CRdiff}
As described in \citetalias{Protheroe2008}, we treat the diffusion of CRs into a molecular cloud described by the above density profiles as analogous to radiative transfer principle of absorption and scattering, giving a cloud an effective optical depth to CRs.
Here the absorption, $\tau_a$ and scattering, $\tau_s$ are:
\begin{equation}
	\tau_a = \int^R_r0.5[2n_{H_2}(r^{\prime})]\sigma_{pp}dr^{\prime}
\end{equation}
and
\begin{equation}
	\tau_s = \int^R_r\frac{c}{3D(E,r^{\prime})}dr^{\prime},
\end{equation}
where $\sigma_{pp}$ is the proton-proton interaction cross-section, and $D(E)$ is the diffusion coefficient as defined in \citet{Gabici2007}:
\begin{equation}
D(E) = 3\times10^{27}\chi\left[\frac{E/(1{\rm~GeV})}{B/(3{\rm~}\mu{\rm G})}\right]^{0.5},
\end{equation}
where $\chi$ is the factor introduced to account for possible suppression of the diffusion of CRs.
This changes the CR intensity at radius $r$, $I_{\rm{\tiny CR}}(E,r)$ by the factor $e^{-\tau_\star(E,r)}$, where $\tau_\star=\tau_a(\tau_a + \tau_s)$.

We briefly note that in order that the synchrotron emission from secondaries be observable, the secondary electrons are required to be produced and to decay within the confines of any putative cloud.
As was shown in \citetalias{Protheroe2008}, the diffusion distance for a magnetic field of 600~\mug in a density of $10^4$~\cmcube is $\sim0.5$~pc, much smaller than the 2.97~pc radius of the cloud.
Thus, given that magnetic field strength obtained for \source is similar, we suggest that it is reasonable to assume that ambient CRs will interact and produce secondaries within the cloud, and not be able to diffuse, or be advected away from the cloud.

Figure~\ref{fig:CRpen} (a) and (b) shows the penetration factor, $e^{-\tau_\star}$, obtained for the density distributions discussed in Section~\ref{sec:density} using a peak density of $2\times10^4$~\cmcube and a magnetic field strength perpendicular to the line of sight, $B_\perp=600$~\mug.
The value of $B_\perp=470$~\mug is motivated by the result of \citet{Johnston2014}, who found a total magnetic field strength of $B\sim470$~\mug, and, following the arguments of \citetalias{Protheroe2008} (i.e., $B_\perp=\pi B_{\textrm{\tiny{LOS}}}/2=\pi B/4$, since $\langle B_{\textrm{\tiny{LOS}}}\rangle=B/2$), hence $B_\perp=470$~\mug is a reasonable estimate.
These values are well within the range of that known in Galactic centre clouds of 120~\mug to 5.7~mG \citep{Johnston2014}.

The importance of Figures~\ref{fig:CRpen}(a) and (b) are that if the conditions are favourable (i.e., few signs of massive star formation, combined with high magnetic fields, densities and masses), then sensitive radio studies of such cloud can illuminate the structures of the clouds; centrally-condensed clouds will exhibit synchrotron emission close to its centre, whereas clouds at later evolutionary stages -- such as evidenced by PDFs with power-law tails and/or inverse-$r$ or Gaussian density profiles -- will contain the bulk of the CRs at larger distances from their centres.
This gives an important link between the evolutionary stage of the cloud and where CRs, and hence synchrotron emission, should be observed.

\begin{figure*}
  \centering
 \includegraphics[width=0.45\textwidth]{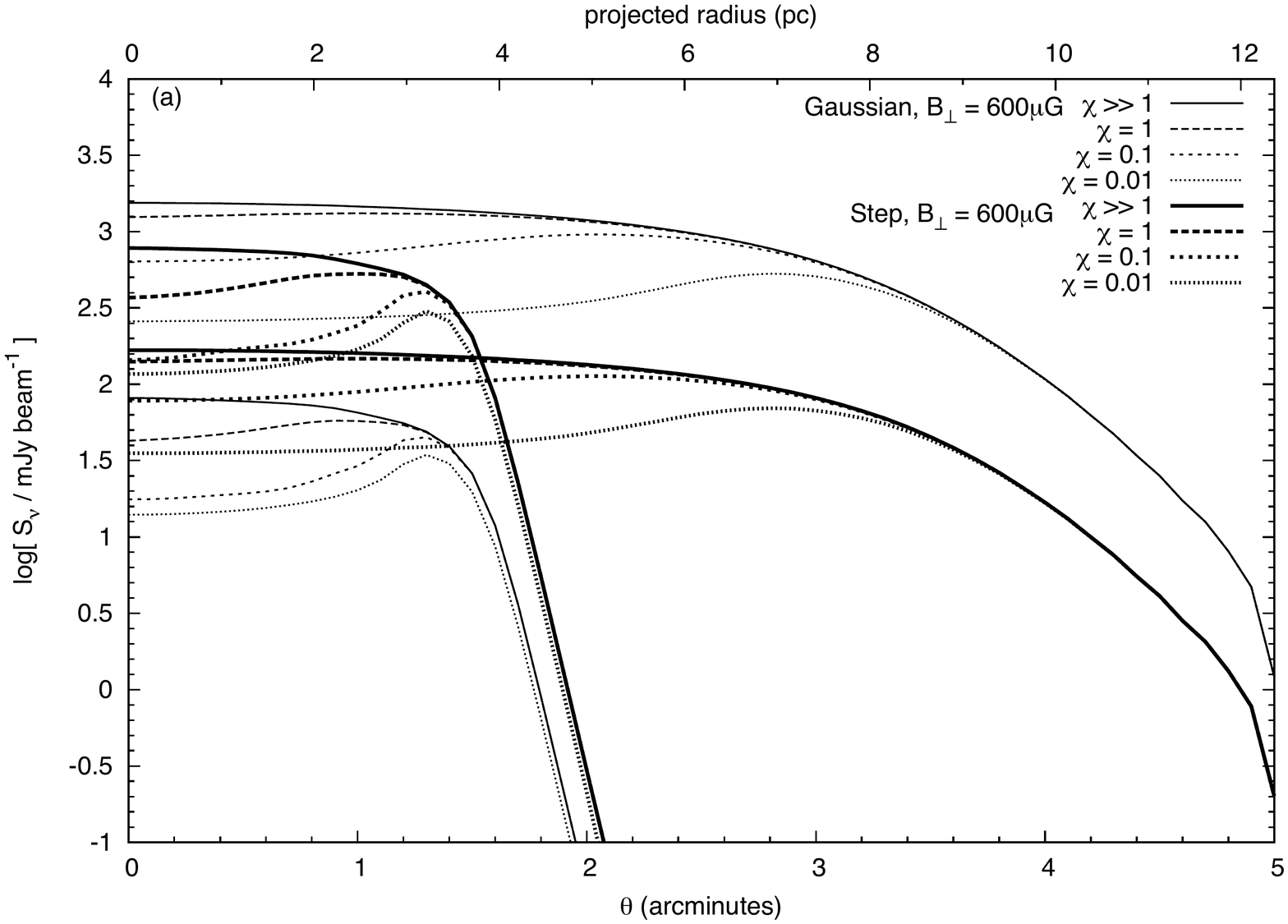}\includegraphics[width=0.45\textwidth]{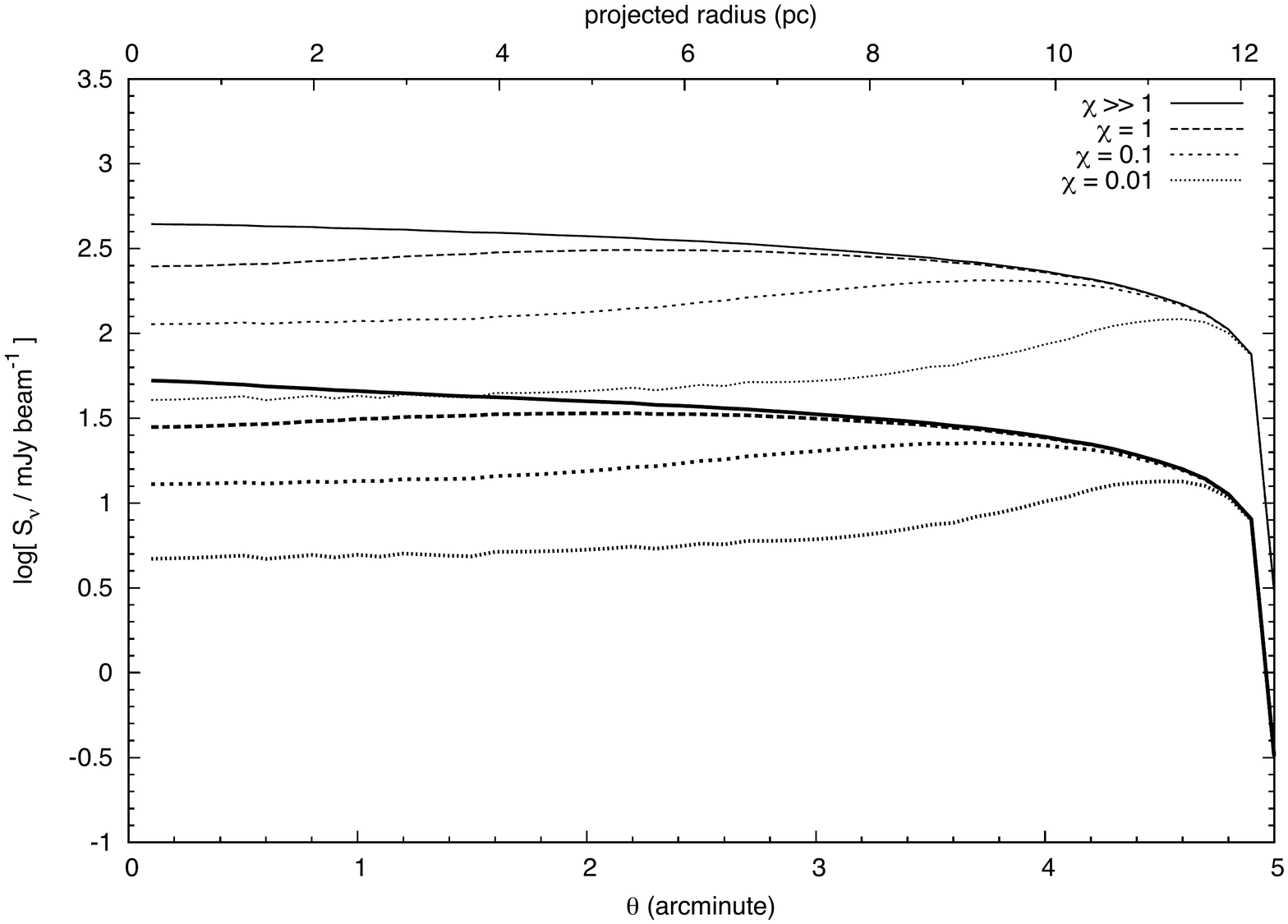}
  \caption{Plots of flux density, $S_\nu$ integrated over a solid angle corresponding to the $2.2''$~VLA beam at 5.3~GHz from \citet{Rodriguez2013}.
  Panel (a) shows the flux density obtained using a Gaussian and step-function density profile for a magnetic field strength of 470 and 600~\mug (thin and thick lines, respectively), for different CR diffusion suppression coefficients, $\chi$, as labelled.
  Panel (b) shows the flux density obtained using $1/r^2$ density profile for the 600 and 470~\mug magnetic field strength (thin and thick lines, respectively).
  We show only the $1/r^2$ density profile here for clarity -- the two profiles are very similar.
  }
  \label{fig:flux}
\end{figure*}

From the synchrotron emissivity and by appropriate integration, the angular distribution of flux density is obtained and shown in Figure~\ref{fig:flux} for a Gaussian and step-function density distribution, in panel (a) and for a $1/r^2$ only (due to them being so similar) density distributions in panel (b).
This figure shows the important role that the overall density distribution plays in the expected synchrotron emissivity as a function of distance from the centre of the cloud.
Because the step-function density distribution reflects a sharp cloud boundary, whereas Gaussian and inverse-$r$ distributions represent a more gradual decline in the density, any resulting synchrotron emission will be observed closer to the centre of the cloud than is expected for the other density distributions (this is also evidenced in Figure~\ref{fig:CRpen}).
And, given that the density distributions of clouds are known to change throughout their life cycles (e.g., the evolution of a power-law tail of PDFs; \citealt{Johnston2014}), synchrotron emission may be observed at different distances from a cloud depending on the stage of (massive) star formation that it is at.
Thus, for a centrally condensed cloud, such as represented as the step-function density profile shown in Figure~\ref{fig:flux} (a), we predict detectable synchrotron emission, on mJy/beam levels at $\sim3-4$~pc from the centre of the cloud.

\section{Discussion and Conclusions}
\subsection{Application to \source}
Figure~\ref{fig:flux} (a) and (b) shows the flux density (in units of Jy/beam, where the beam is $2.2''$) expected for a cloud with the characteristics of \source.
It is significant then that in their radio continuum study of \source, \citet{Rodriguez2013} found three compact, thermal sources but also three compact sources that possess a non-thermal spectrum between 5.3 and 20.9~GHz (i.e., sources JVLA~1, 2, 6 and 7 -- see their Figure~1).
Figure~\ref{fig:flux} shows that for reasonable magnetic field values, our model can explain the compact, non-thermal sources observed towards \source, {\it assuming that they are related to the clump}, and are not background galaxies.
There are, however, good reasons to think that at least two of these non-thermal sources -- JVLA~6 and 1 -- are indeed associated with \source, whilst the other two -- JVLA~2 and 7 are background sources.
Firstly, sources JVLA~2 and 7 are located further away from the centre of the source than JVLA~6 and 1, for which JVLA~6 is located within the blue contours defining the source in Figure~1 of \citet{Rodriguez2013}, and JVLA~6 lying (in projection) just exterior to the source at the position of the greatest density gradient of the clump.
This fits well with the scenario outlined in the previous section; synchrotron emission located at the edge of the cloud.
Secondly, the JVLA~6 and 1 also possess flatter spectral indices (viz. $\alpha=-0.3$ and --0.9 for JLVA source 1 and 6, for a spectral index calculated as $\alpha=d\log S_\nu/d\log\nu$), than JVLA~2 and 7, suggesting that the latter are indeed background sources.

If sources JVLA~1 and 6 are co-located with \source, Figure~\ref{fig:flux} suggests that in their environment, the CR diffusion suppression factor is $\sim0.1$ and 0.01, respectively, with a magnetic field value of $\sim470$~\mug.
We obtain these estimates by comparing the distance of these sources away from the centre of the source, and matching the flux density observed at 5.3~GHz, with that predicted by our model.
This conclusion is entirely reasonable, given the parameters in our model, and may even be an underestimate since given the density of CR sources is expected to rise towards the GC, one might expect an increase in the flux of CRs there.
We also note that if the density distribution of a molecular cloud changes as it evolves, then so will the location of synchrotron emission resulting from secondary electrons.
If a step-function-like density distribution is typical of molecular clouds that are just starting to form massive stars, then any synchrotron emission would be observed near to the clouds centre, whereas, if an inverse-$r$ function is more typical of the density structure, then any such emission would be located further from the cloud centre.
Finally, we note that the above applies only to the overall density of the cloud, and not to fine structures.
However, if, as shown in \cite{Battersby2014}, massive star formation results in cores with a mean ambient density structure of $\rho\propto r^{-1.8}$, then the sensitive new radio telescopes, such as JVLA, LOFAR, ASKAP and the SKA should be able to observe non-thermal sources which appear to be background sources shining through the cloud, but may in fact be compact regions of non-thermal synchrotron emission due to secondary electrons due to small, condensed cores within the larger parent cloud.
Finally, given the additional sensitivity of the next generation of radio telescopes, it is possible that the broad emission, and not just the limb-brightened regions could be detected.

\section*{Acknowledgments} This research has made use of the SIMBAD database, operated at CDS, Strasbourg, France.
We thank the anonymous referee for comments that greatly improved the manuscript.
We thank S. Thoudam, C. van Eck, M. Haverkorn, J. Ott for enlightening discussions, and R. Yang and E. Kafexhiu for discussions about the production spectrum of secondary electrons and positrons.

\bibliographystyle{apj}
\bibliography{bibliography}

\end{document}